\pdfoutput=1
\documentclass[a4paper,american,floatfix,pdftex,superscriptaddress,twoside,%
journal=jacsat,achemso,%
citeautoscript,
reprint,twocolumn
]{revtex4-2}%
\usepackage{amsfonts,amsmath,amssymb}
\usepackage[T1]{fontenc}
\usepackage{graphicx}%
\usepackage[utf8]{inputenc}
\usepackage{hypernat}
\usepackage[ams,todos]{CMImacros}

\graphicspath{{./fig/}{./}} 
\hypersetup{citebordercolor=green,linkbordercolor=red,urlbordercolor=blue}

\newcommand*{\rodsize}[2]{\ensuremath{(#1,~#2)}}

\newcommand{\cfeldesy}{\affiliation{Center for Free-Electron Laser Science CFEL, Deutsches
      Elektronen-Synchrotron DESY, Notkestr. 85, 22607 Hamburg, Germany}}%
\newcommand{\ucg}{\affiliation{Department of Sciences, University College Groningen, University of
      Groningen, 9718 BG, Groningen, Netherlands}}%
\newcommand{\ucgr}{\affiliation{Laboratory of Computational Biology, National Heart, Lung and Blood
      Institute, National Institutes of Health, Bethesda, Maryland 20892, USA}}%
\newcommand{\snrs}{\affiliation{Laboratoire de Météorologie Dynamique/IPSL, CNRS, Ecole
      polytechnique, Institut Polytechnique de Paris, Sorbonne Université, Ecole Normale Supérieure,
      Université PSL, 91120 Palaiseau, France}}%
\newcommand{\uhhcui}{\affiliation{Center for Ultrafast Imaging, Universität Hamburg, Luruper
      Chaussee 149, 22761 Hamburg, Germany}}%
\newcommand{\uhhphys}{\affiliation{Department of Physics, Universität Hamburg, Luruper Chaussee 149,
      22761 Hamburg, Germany}}%
\newcommand{\jkemail}{\email[Email:~]{jochen.kuepper@cfel.de}}%
\newcommand{\cmiweb}{\homepage[website:~]{https://www.controlled-molecule-imaging.org}}%

\begin{document}
\title{Laser-induced alignment of nanoparticles and macromolecules \\
   for coherent-diffractive-imaging applications}%
\author{Muhamed Amin}\cfeldesy\ucgr\ucg%
\author{Jean-Michel\ Hartmann}\snrs%
\author{Amit K.\ Samanta}\cfeldesy\uhhcui%
\author{Jochen Küpper}\jkemail\cmiweb\cfeldesy\uhhcui\uhhphys%
\date{\today}%
\begin{abstract}\noindent
   \textbf{Abstract} Laser-induced alignment of particles and molecules was long envisioned to support
   three-dimensional structure determination using ``single-molecule diffraction'' with x-ray
   free-electron lasers [PRL 92, 198102 (2004)]. However, the alignment of isolated macromolecules
   has not yet been demonstrated, also because quantitative modeling is very expensive. We
   computationally demonstrated that the alignment of nanorods and proteins is possible with
   standard laser technology. We performed a comprehensive analysis on the dependence of the degree
   of alignment on molecular properties and experimental details, \eg, particle temperature and
   laser-pulse energy. Considering the polarizability anisotropy of about 150,000 proteins, our
   analysis revealed that most of these proteins can be aligned using realistic experimental
   parameters.
\end{abstract}
\maketitle



\section{Introduction}
X-ray free-electron lasers (XFELs) promise the diffractive imaging of macromolecules and
nanoparticles at atomic resolution~\cite{Neutze:Nature406:752} and the recording of ``molecular
movies'' of their structural dynamics~\cite{Barty:ARPC64:415}. Laser alignment was
long-envisioned~\cite{Spence:PRL92:198102} to maximize the information retrieval from the recorded
images.

In these experiments high-intensity, femtosecond, x-ray pulses interact with individual molecules
and yield diffraction patterns using the ``diffraction-before-destruction''
approach~\cite{Neutze:Nature406:752, Spence:PRL92:198102, Chapman:NatMater8:299,
   Bogan:NanoLett8:310, Barty:ARPC64:415}. In single-particle imaging
(SPI)~\cite{Spence:PRL92:198102, Spence:SpringerHandbook:SPI:2019} a large set of diffraction
patterns of individual molecules is collected. These two-dimensional images, \eg, from randomly
oriented samples, would then be computationally assembled to a diffraction volume to retrieve the
three-dimensional structure~\cite{Barty:ARPC64:415}. This was achieved, albeit not to atomic
resolution, for nanoparticles~\cite{Seibert:Nature470:78, Ayyer:Optica8:15, Hoeing:NanoLett23:5943}.

In ``standard SPI'' the images do not contain \emph{a priori} information of the molecules'
orientation and, so far, this uncertainty is attacked \emph{in silico}~\cite{Ayyer:Optica8:15}.
Significant efforts were made in improving the reconstruction process and the achievable resolution,
but it is still a highly challenging task, especially for weakly scattering molecules, \eg,
individual proteins, where diffraction signals from single molecules are generally not sufficient to
allow for the computational averaging and sorting~\cite{Ayyer:Nature530:202, Ayyer:Optica7:593}.
This is one of the major bottlenecks for atomic-spatial-resolution SPI.

As originally proposed by Spence \etal~\cite{Spence:PRL92:198102}, imaging molecular samples with
controlled alignment or orientation~\cite{note:alignment+orientation} significantly mitigates this
problem by allowing to sum the diffraction signals from many identically-aligned molecules to
provide a much stronger signal, thus improving the reconstruction step and paving the way toward
atomic-resolution SPI~\cite{Spence:PRL92:198102, Filsinger:PCCP13:2076, Barty:ARPC64:415}. Careful
analysis of simulated diffraction patterns of laser-aligned proteins demonstrated that it is
possible to observe their secondary structure with only reasonably-strong degrees of alignment
$\costhreeD\geq0.9$~\cite{Spence:ActaCrystA61:237}.

The alignment of small molecules using external electric fields, including strong dc fields and
optical fields from moderately intense, nonresonant light pulses was studied
extensively~\cite{Friedrich:Nature353:412, Block:PRL68:1303, RoscaPruna:PRL87:153902,
   Larsen:JCP111:7774, Stapelfeldt:RMP75:543, Trippel:PRA89:051401R, Koch:RMP91:035005}. Laser
alignment or mixed-field orientation does allow for three-dimensional
confinement~\cite{Larsen:PRL85:2470, Holmegaard:NatPhys6:428}, whereas dc-field brute-force
orientation only enables one-dimensional confinement. Strong alignment was achieved for linear,
symmetric top, and asymmetric top molecules in the adiabatic~\cite{Larsen:JCP111:7774,
   Holmegaard:PRL102:023001}, intermediate~\cite{Trippel:MP111:1738, Trippel:PRA89:051401R}, and
impulsive~\cite{RoscaPruna:PRL87:153902, Karamatskos:NatComm10:3364} regimes. Considerable efforts
were made for laser-induced alignment of ``large'' molecules~\cite{Mullins:NatComm13:1431,
   Pentlehner:PRA87:063401} including complex, floppy polyatomic
molecules~\cite{Chatterley:NatCommun10:133} and weakly bound molecular
complexes~\cite{Trippel:JCP148:101103}. Such aligned-molecules samples were studied by
electron~\cite{Park:ACIE47:9496, Hensley:PRL109:133202} and x-ray diffractive
imaging~\cite{Kuepper:PRL112:083002}. Possibilities to laser-align large biomolecules without
deterioration of the secondary structure were proposed~\cite{Spence:PRL92:198102, Barty:ARPC64:415},
but no alignment for such systems was reported.

For macromolecules, there are several challenges for achieving the required alignment. Theoretical
predictions supported by \emph{in silico} analysis are a very important step to guide experiments.
However, atomistic molecular dynamics (MD) simulations are computationally very expensive for large
particles, especially for nanosecond timescales~\cite{Noe:BPJ108:228}. In addition, to accurately
predict ensemble-averaged single-particle diffraction patterns, the simulation of a large
distribution of particles is essential. Ensemble computations are also crucial for studying
temperature effects, which are important both for laser-alignment control and for preserving the
secondary structures of macromolecules.

Here, we predicted and analyzed the laser-induced alignment of nanoparticles and biomolecules. These
were treated as rigid bodies, supported by previous molecular-dynamics simulations of structural
changes in strong electric fields~\cite{Marklund:JPCL8:4540, Brodmerkel:ProteinJ42:205}. The key
parameters are the overall polarizability, shape, and their anisotropies, the temperature, and the
alignment-laser field. We disentangled how these parameters can be tuned for maximum alignment of
metal-nanorods and how this can be exploited for the strong alignment of biological macromolecules,
\eg, proteins. Overall, our computations demonstrate that most proteins can be strongly laser
aligned.

\section{Computational methods}
\label{sec:methods}
The response of the particles to a nonresonant electric field was calculated based on their
polarizability tensors, which yielded the time-dependent induced dipole moments. For metallic
nanorods, the polarizabilities were directly obtained by solving Laplace’s equation with Dirichlet
boundary conditions and using Monte Carlo path integral methods~\cite{Juba:JRNIST122:20}. For
proteins, the polarizabilities were derived from the same calculations using regression-based
scaling~\cite{Amin:JPCL10:2938, Amin:ChemistryOpen9:691}.

Rotational dynamics of the particles were calculated classically~\cite{Ma:PRR3:023192}. Molecular
ensembles were set up with random initial orientations and initial angular velocities according to a
Boltzmann distribution at the given temperature. Each run included 20,000 particles. The angular
positions were stored in quaternions throughout the calculation. The inertial tensors of the
artificial nanorods were calculated for cylinders. For proteins they were calculated based on the
atomic masses and coordinates from the protein data bank (PDB)~\cite{Berman:NuclAcidRes28:235}.

Electric fields of the laser pulses were represented by Gaussian functions with variable peak
intensities. We used a temporal full-width at half maximum (FWHM) of 8~ns corresponding to standard
Q-switched Nd:YAG lasers (1064~nm). We also assumed linearly-polarized laser pulses with intensities
of $10^{10}\ldots10^{12}$~\Wpcmcm.

The particles' phase-space positions were propagated in time by integrating Euler's equations using
a new, Python-based, openly available software package CMIclassirot~\cite{Amin:CMIclassirot}, based
on and checked against previous classical-alignment computations~\cite{Hartmann:JCP136:184302}. All
simulations propagated the particles for 50~ns; this was extended to 200~ns for the data in
\autoref{fig:nanorods}.

The effect of resonances was ignored, \ie, the laser-field frequency was assumed to be far off
resonance from the molecules' and nanoparticles' absorption. For metal nanoparticles, localized
surface-plasmon resonances occur at specific wavelengths, which strongly depend on the
nanoparticles' size, shape, and material~\cite{Kooij:PCCP8:3349} and the surrounding medium. This
effect can be avoided by an appropriate choice of the laser wavelength~\cite{Kooij:PCCP8:3349}.

\section{Results}
\begin{figure}[b]
   \includegraphics[width=\linewidth]{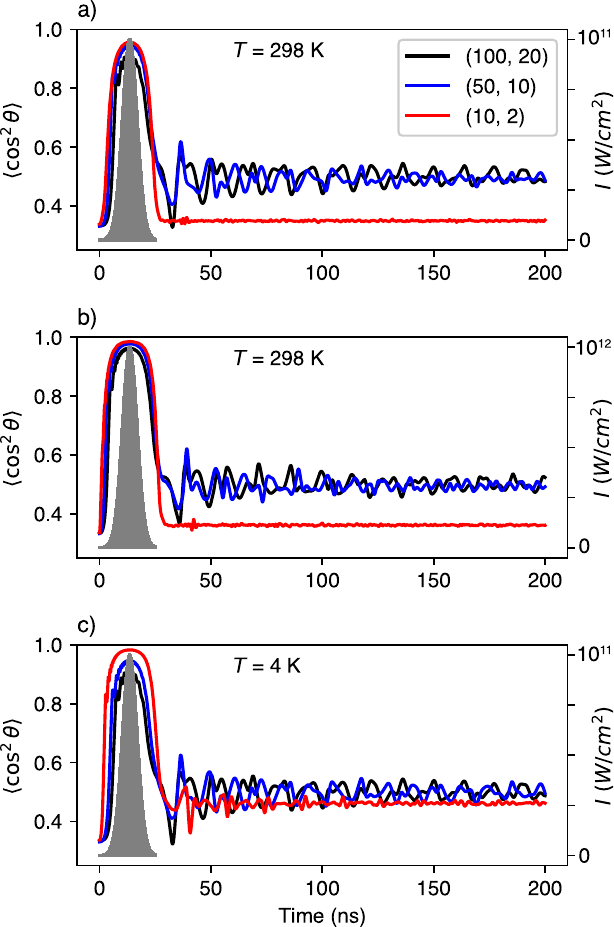}%
   \caption{Degree of alignment of different-sized gold nanorods obtained for selected temperatures
      and laser intensities: a)~\mbox{$T=298$~K}, $I_{max}=10^{11}~\Wpcmcm$; b)~$T=298$~K,
      \mbox{$I_{max}=10^{12}~\Wpcmcm$}; c)~$T=4$~K, $I_{max}=10^{11}~\Wpcmcm$. The temporal laser
      profile is indicated by the shaded gray area and the corresponding field intensities are
      specified on the secondary axes. The polarizability components of the three particles are
      proportional to the particles' volumes~\cite{Amin:JPCL10:2938}, specified in the legend, and
      are $\alpha\approx0.24V$ ($7540~\text{nm}^3$, $942~\text{nm}^3$, $7.5~\text{nm}^3$),
      $\alpha_\parallel\approx0.58V$, $\alpha_\perp\approx0.07V$.}
   \label{fig:nanorods}
\end{figure}

The computed time-dependent degrees of laser-induced alignment of different nanorods are shown in
\autoref{fig:nanorods} for different temperatures ($T$) and laser intensities ($I$). Throughout this
manuscript, nanorod sizes are represented as ($\text{length}$/nm, $\text{diameter}$/nm), with the
calculations performed for \rodsize{10}{2}, \rodsize{50}{10}, \rodsize{100}{20}.
During the pulse and with particles initially at room temperature, the degree of alignment for all
three particles, \autoref[a,~b]{fig:nanorods}, follows the temporal laser profile on the rising edge
quasi adiabatically~\cite{Trippel:PRA89:051401R}. The smallest particle \rodsize{10}{2} exhibits the
largest alignment of $\costhreeD=0.96$.

The laser turn-off dynamics show considerable differences: The two larger nanorods exhibit permanent
alignment after the laser pulse, whereas the small nanorod quasi-adiabatically follows the temporal
laser profile to an isotropic field-free angular distribution $\costhreeD=1/3$. This can be
rationalized by comparing the rotation periods, \ie, the temperature-dependent average time needed
for each nanorod to rotate around its center of mass by \degree{360}, to the laser-pulse duration.
For the largest nanorods at room temperature, the rotational periods are $\ordsim10~\us$
(\mbox{$\nu\approx100$~kHz}, $\omega\approx6.3\cdot10^{5}$~rad/s), three orders of magnitude larger
than the pulse duration. These particles exhibit non-adiabatic dynamics with permanent alignment
after the pulse is off, \autoref[a,~b]{fig:nanorods}. Already at low intensity, $I=10^{11}~\Wpcmcm$,
these particles are confined to rotate in a plane containing the laser polarization vector,
corresponding to $\costhreeD=0.5$, \autoref[a]{fig:nanorods}. No increase in the permanent alignment
is possible at $I=10^{12}~\Wpcmcm$, \autoref[b]{fig:nanorods}. The small nanorod has a rotational
period of $\ordsim20$~ns, comparable to the laser pulse duration, and a quasi-adiabatic response
without permanent alignment is observed~\cite{Trippel:MP111:1738, Trippel:PRA89:051401R},
\autoref[a,~b]{fig:nanorods}.

At 4~K~\cite{Samanta:StructDyn7:024304}, the rotational period of the \rodsize{10}{2} rod increases
to 400~ns, resulting in the transition to the non-adiabatic regime and field-free permanent
alignment is observed, \autoref[c]{fig:nanorods}.

For the larger nanorods (blue, black) the degree of permanent alignment after the pulse shows a
strong oscillation that decays with time due to the progressive dephasing between the rotations of
the confined nanorods, which start to rotate with different angular velocities, but significantly
slower than for the small particles. Furthermore, to ensure that these oscillations were not a
result of undersampling, we tested convergence for particles numbers up to 200,000 and obtained
the same oscillations.

While our simulations take into account the full polarizability tensor, it is instructive to study
the degree of alignment as a function of the nanorod shape described by the ratio of its principal
moments of polarizability, which correlates with the polarizability anisotropy:
\begin{align}
  \label{eqn:pol-anisotropy}
  \alpha_\parallel &= \max\left(\alpha_{11},\alpha_{22},\alpha_{33}\right) \notag \\
  \alpha_\perp &= \left(3\alpha-\alpha_\parallel\right)/2 \\
  \alpha_r &= \alpha_\parallel / \alpha_\perp \notag
\end{align}
$\alpha_{11}$, $\alpha_{22}$ and $\alpha_{33}$ are the principal moments of polarizability and
$\alpha$ denotes their average. For $\alpha_r$ the scaling factor
applied to the elements of the polarizability tensors to account for the dielectric medium of
proteins~\cite{Amin:JPCL10:2938} cancels out, and does not have to be included in this discussion of
shape.

\begin{figure}
   \includegraphics[width=\linewidth]{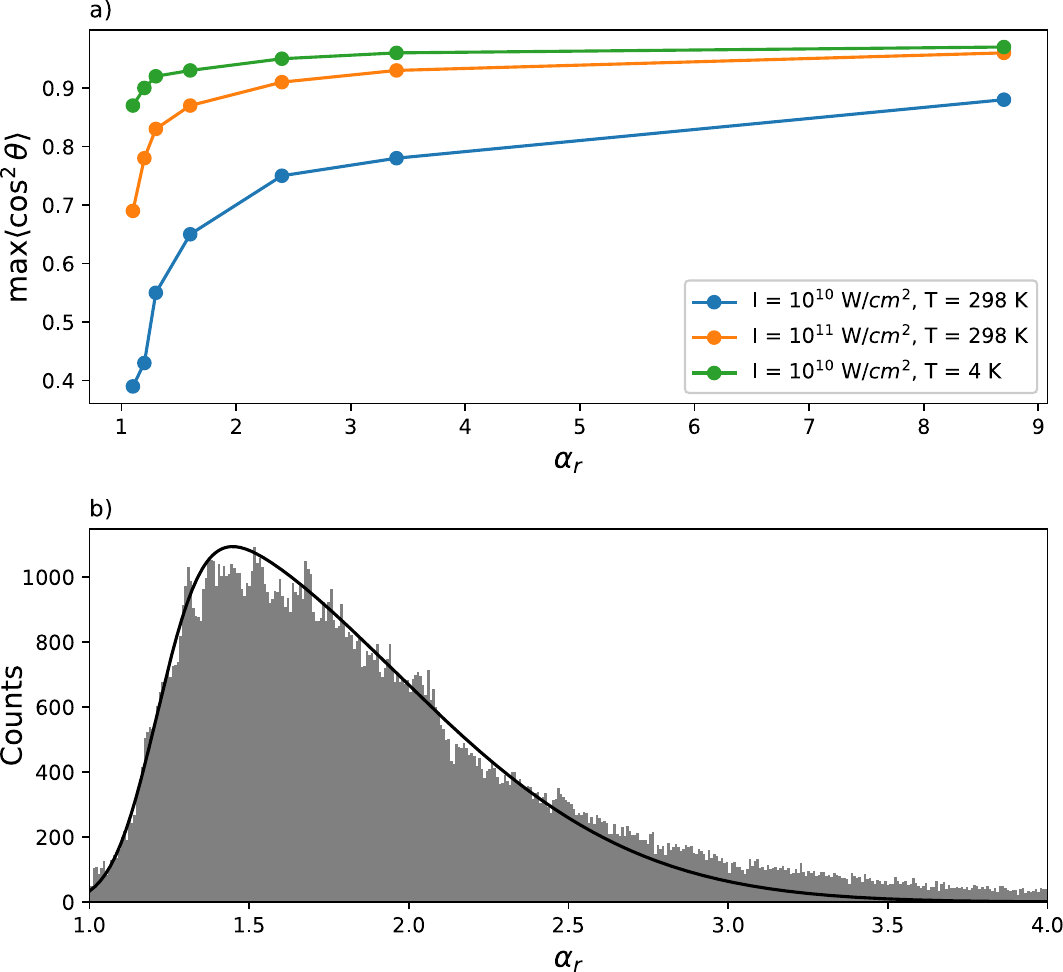}%
   \caption{a) Maximum degree of alignment, $\max\costhreeD$, as a function of the polarizability
      ratio $\alpha_r$ at different temperatures and intensities for a nonorod of
      $V=10\pi~\text{nm}^3$, \ie, $\alpha\approx7.5~\text{nm}^3$. At low temperature a high degree
      of alignment is achieved even at very low intensity. b) The $\alpha_r$ distribution of about
      150,000 proteins in the protein data bank (PDB) approximately follows a skew-normal
      distribution with a location of $1.2$, scale of $0.7$ and skewness parameter of $6.5$.}
   \label{fig:PDB}
\end{figure}
\autoref[a]{fig:PDB} shows the dependence of the maximum degree of alignment $\max\costhreeD$ on
$\alpha_r$ for $I=10^{10}~\Wpcmcm$ and $I=10^{11}~\Wpcmcm$. There is a quick rise of
$\max\costhreeD$ for $\alpha_r$ in the range $1.2\ldots2.5$, depending on temperature, confirming
that with increasing particle anisotropy there is a significant increase in the maximum degree of
alignment. Further increasing $\alpha_r$ led to a slow increase in the maximum alignment toward an
asymptotic maximum. For higher intensity one observes stronger alignment, especially for small
values of $\alpha_r$. The same holds for lower temperatures, which enable significantly increased
alignment even at lower laser intensities~\cite{Kumarappan:JCP125:194309, Holmegaard:PRL102:023001}.
Specifically, for $10^{10}~\Wpcmcm$ and at room-temperature it requires $\alpha_r>8$ to obtain a
reasonable degree of alignment of $\costhreeD>0.8$, whereas at 4~K $\costhreeD=0.9$ is achieved at
$\alpha_r=1.5$ .

Proteins are a principal target for SPI. Thus, we calculated the polarizability ratio $\alpha_r$ of
about 150,000 proteins using a database we built previously~\cite{Amin:ChemistryOpen9:691} using
ZENO~\cite{Juba:JRNIST122:20, Amin:ChemistryOpen9:691} on the proteins' PDB structures, see
\autoref[b]{fig:PDB}. The tensors were diagonalized to put all molecules in the polarizability
frame, their $\alpha_r$ values were computed according to \eqref{eqn:pol-anisotropy}, and summarized
in the histogram in \autoref[b]{fig:PDB}. $\alpha_r$ follows a skew-normal distribution with a
location of $\ordsim1.2$. As $96~\%$ of the data have $\alpha_r>1.2$, this indicates that a
significant fraction of these proteins has sufficiently anisotropic polarizabilities to be strongly
aligned. The polarizability volumes of this set of proteins follow a skewed-normal distribution with
a location parameter equal to 3.3~nm$^3$~\cite{Amin:ChemistryOpen9:691}, showing the similarity with
the small nanorod \rodsize{10}{2} with $\alpha=7.5~\text{nm}^3$.

\begin{figure}
   \includegraphics[width=\linewidth]{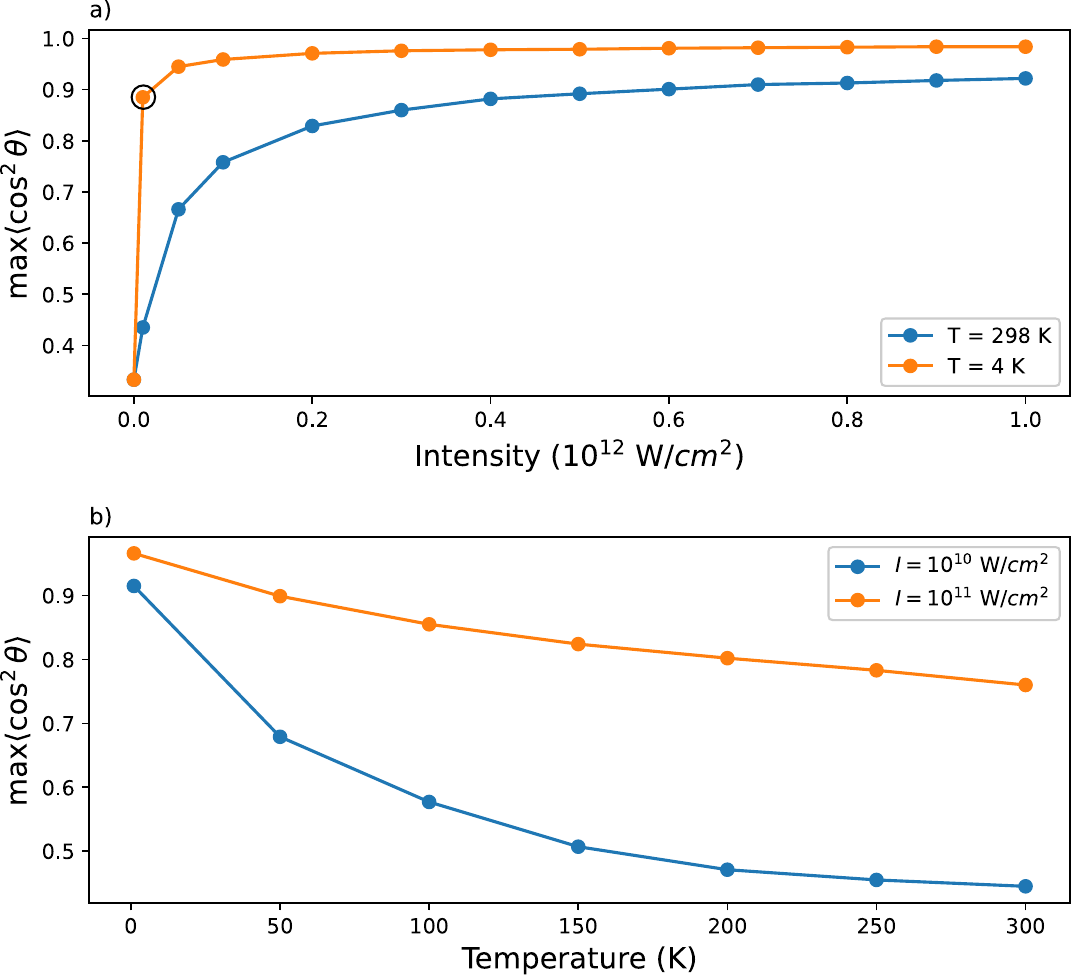}%
   \caption{a) The effect of laser intensity on the degree of alignment of a gold nanorod with
      $\alpha_r=1.2$ and size \rodsize{3.6}{3.2} at 298~K (blue) and 4~K (orange). b) The effect of
      temperature on the degree of alignment of the same gold nanorod for two different laser
      intensities, $10^{10}~\Wpcmcm$ (blue) and $10^{11}~\Wpcmcm$ (orange). }
   \label{fig:Int}
\end{figure}

To provide further understanding of the effect of the shape of the object on the maximum achieved
alignment, we studied the effect of laser intensity and particle temperature for a gold nanorod with
$\alpha_r=1.2$ and size \rodsize{3.6}{3.2}, see \autoref[a]{fig:Int}. The degree of alignment
quickly increases at low intensities, depending on temperature, and saturates toward the asymptotic
limit, \emph{vide supra}. Again, the achievable alignment at 4~K is significantly higher than at
room temperature. In the latter case, $\max\costhreeD=0.9$ is achieved for the highest intensity of
$10^{12}~\Wpcmcm$, whereas a 4~K sample enables very strong alignment of $\max\costhreeD\approx0.9$
at $10^{10}~\Wpcmcm$ (gray circle) and $\max\costhreeD=0.95$ at $5\cdot10^{10}~\Wpcmcm$. Similarly,
the results in \autoref[b]{fig:Int} demonstrate the decrease of alignment with increasing
temperature, especially for lower laser intensities.

Thus, based on our model system of a nanorod with an anisotropy similar to that of most proteins,
\autoref{fig:PDB}, strong alignment is achievable for most proteins at 298~K using 10~ns pulses with
peak intensities around $10^{11}~\Wpcmcm$. This alignment can be improved significantly by
exploiting cryogenically cooled ($\smaller4$~K)~\cite{Samanta:StructDyn7:024304} proteins even with
a much weaker laser pulse. Furthermore, the cooling will also reduce the chance of structural damage
by intense fields, \emph{vide infra}.

\begin{figure}
   \includegraphics[width=\linewidth]{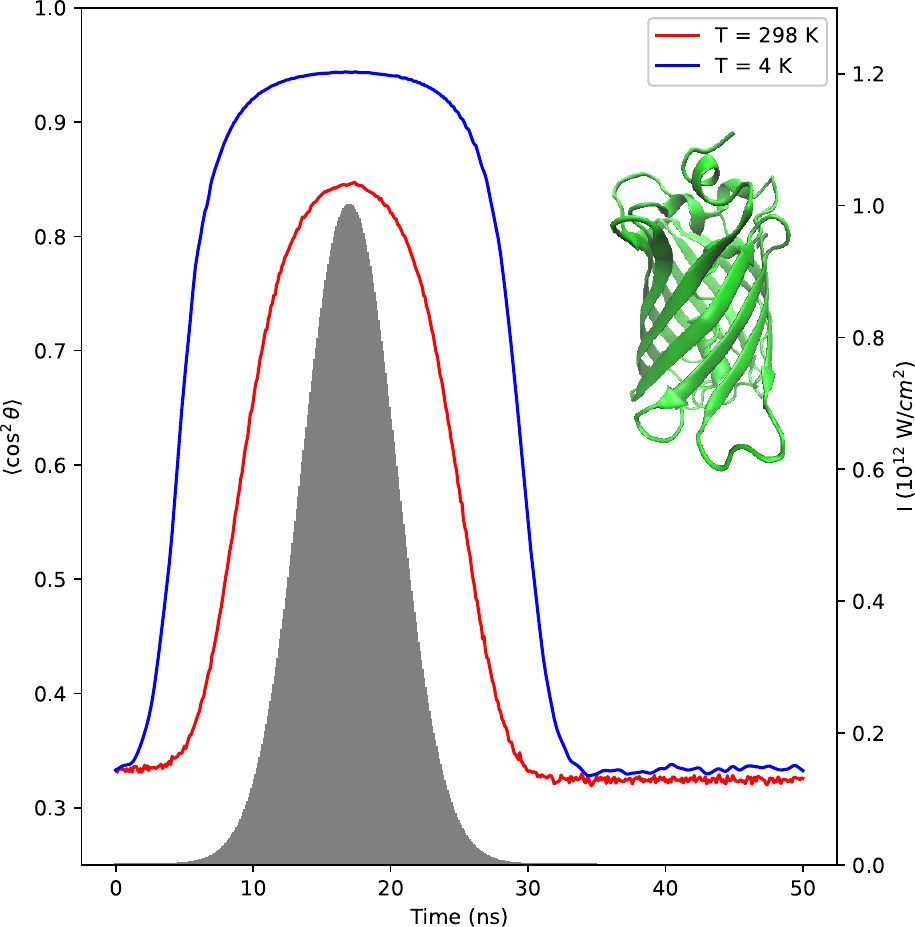}%
   \caption{Degree of alignment of the green fluorescent protein, depicted in the inset, at 298~K
      (red) and 4~K (blue) using $10^{12}~\Wpcmcm$ pulse intensity (shaded area in gray). The
      protein has a cylindrical shape with a base diameter of $2.5~\text{nm}$ and a length of
      $5.3~\text{nm}$. }
   \label{fig:protein}
\end{figure}
To simulate the alignment of an actual protein, we applied our method to the prototypical green
fluorescent protein (GFP), which has a cylindrical shape ($\alpha_r=1.5$) and a volume of
26~$\text{nm}^3$, comparable to most proteins in the PDB database~\cite{Amin:ChemistryOpen9:691},
see \autoref{fig:PDB}. We calculated the inertial and polarizability tensors based on the PDB
structure 1GFL~\cite{Yang:NatBio14:1246}. The perfect-conductor polarizability tensors were scaled
by a factor of $0.4$ to account for the dielectric properties of proteins,
$\epsilon_r=3.2$~\cite{Amin:ChemistryOpen9:691}. As shown in \autoref{fig:protein}, moderate
alignment of $\max\costhreeD=0.84$ was obtained at room temperature for a peak intensity of
$10^{12}~\Wpcmcm$. However, at 4~K $\max\costhreeD=0.85$ was already achieved for
$2\cdot10^{10}~\Wpcmcm$ and for $10^{12}~\Wpcmcm$ we obtained $\max\costhreeD=0.94$.

The original conceptional proposal~\cite{Spence:PRL92:198102} suggested to use cooling by helium
nanodroplets to reach temperatures of $\ordsim0.4$~K. This cooling method was successfully applied
to small molecules~\cite{Hartmann:Science272:1631, Choi:IRPC25:15} and small
proteins~\cite{Bierau:PRL105:133402, Alghamdi:JPCA121:6671}. Diffraction off atoms and molecules
embedded in helium droplets was also recorded~\cite{Gomez:Science345:906, Zhang:JPCL5:1801}, but for
weak-scattering SPI of, \eg, single proteins, the helium background signal poses a challenge.
Comparable temperatures can be achieved for gas-phase molecules through buffer-gas
cooling~\cite{Hutzler:CR112:4803, Samanta:StructDyn7:024304}, down to $\smaller10$~mK using dilution
refrigerators~\cite{Weinstein:Nature395:148}. Decreasing the temperature from 4~K to, e.g., 400~mK,
would allow strong alignment with even weaker laser pulses and thus further reduce the chance of
structural damage. For GFP at 400~mK $\max\costhreeD=0.92$ would be achieved for a peak intensity of
$2\cdot10^{10}~\Wpcmcm$.

In principle, non-resonant fields inducing alignment could also couple low-frequency vibrational
modes of the particles through stimulated inelastic Raman transitions. For significant excitation,
this could distort the molecules' geometries, which can be avoided as the bandwidth of the laser can
be very narrow: The Fourier-transform limit of a 10~ns pulse is $\ordsim10^{-3}$~\invcm and even
monochromatic light could be utilized~\cite{Deppe:OptExp23:28491}. For macromolecules, this spectral
bandwidth does not overlap with even the lowest-wavenumber excitations, which are typically
$\gtrsim1~\invcm$~\cite{Rischel:PNAS95:12306, Ford:PRE67:051924, Bocko:ProcEng96:21}. Thus no
stimulated, or only very weak non-resonant spontaneous, Raman scattering and correspondingly small
structural changes are expected. Moreover, the time-scales for such motions, if excited and even if
corresponding to $\ordsim10^{-3}$~\invcm, would be correspondingly slow, \eg, nanoseconds. This
analysis is in line with MD simulations for proteins in time-dependent dc fields that were stronger
than those considered here, where relevant structural changes still occur only on nanosecond
times~\cite{Marklund:JPCL8:4540, Sinelnikova:BPJ120:3709} as well as with experimental
investigations resulting in negligible structural damage through resonant-Raman vibrational
excitations in proteins~\cite{Maugeria:JACS140:1471}.

Considering the large polarizabilities of the studied molecules and an inverse
polarizability–ionization-energy relationship in small molecules~\cite{Brinck:JCP98:4305},
ionization by the laser field could be another damage mechanism. However, the ionization energies
\Ei of compound systems such as proteins do not scale correspondingly. The polarizabilities are
directly due to the orders-of-magnitudes increased length, \ie, the excursion of electron density
over the whole molecules. In contrast, ionization is a local event where an electron is removed from
a single aromatic amino acids, \eg, indole. This is supported by computed \Ei{s} of the amino acid
tryptophan (Trp) and two small proteins, Trp-cage (PDB ID:1LY2) and FSD-EY (PDB ID:1FME), yielding
$\Ei=7.51, 5.51, 5.39$~eV, respectively, in comparison to their polarizability volumes of
$0.022, 0.221, 0.391~\text{nm}^3$. We also calculated the Hirshfeld charges for Trp-cage and FSD-EY,
which showed that the electron density is mostly perturbed in the indole chromophores. Thus, the
moderate field strengths necessary for alignment, particularly for cold samples, should not lead to
any significant ionization of the molecules.

\section{Conclusions}
In summary, we computationally demonstrated strong laser-induced alignment of isolated nanoparticles
and biological macromolecules for realistic experimental capabilities using simulations based on the
classical dynamics of rigid bodies. We sampled the thermal initial phase space and the following
dynamics over tens of nanoseconds of tens of thousands of particles. We disentangled the dependence
of the degree of alignment on laser intensity, sample temperature, and the molecule's size and
polarizability. A very high degree of alignment can be achieved for cryogenically-cooled proteins at
a moderate laser intensity of $10^{10}~\Wpcmcm$, which should not cause radiation damage. This high
degree of control, $\costhreeD\geq0.94$, paves the way for future atomic-resolution single-molecule
x-ray and electron diffractive imaging experiments. Furthermore, the foreseen atomic spatial and
femtosecond temporal resolution of such experiments provide the prerequisites for future time
resolved studies of ultrafast biochemical dynamics.

Our approach provides clear insight into the optical control of macromolecules and a valuable tool
for exploring the experimental parameters for successful laser-induced alignment. Envisioned future
experiments plan to make use of our cryogenic cooling setup~\cite{Samanta:StructDyn7:024304}
together with efficient laser control to achieve a very high degree of alignment for shock-frozen
proteins and, in turn, sub-nanometer resolution in single particle x-ray imaging.
This framework will prove useful for furthering the field of single-particle x-ray imaging and would
allow us to observe atomically resolved snapshots of ultrafast chemical dynamics.

The achievable strong laser alignment of nanoscopic objects could have further applications in
nanoscience~\cite{Krajewski:Nanoscale:9:16511} as well as in nanoscale quantum optics or quantum
sensing~\cite{Stickler:NatRevPhys3:589}.

\section*{Acknowledgment}
We acknowledge financial support by Deutsches Elektronen-Synchrotron DESY, a member of the Helmholtz
Association (HGF) and the use of the Maxwell computational resources operated at DESY. This work was
supported by the European Research Council under the European Union’s Seventh Framework Program
(FP7/2007-2013) through the Consolidator Grant COMOTION (614507) and the Cluster of Excellence
``Advanced Imaging of Matter'' (AIM, EXC 2056, ID 390715994) of the Deutsche Forschungsgemeinschaft
(DFG).

\section{Code Availability}
Classical rotational dynamics simulations were performed using CMIclassrot, available at
\url{https://github.com/CFEL-CMI/CMIclassirot.git}.

\bibliography{string,cmi}%

\providecommand{\latin}[1]{#1}
\makeatletter
\providecommand{\doi}
  {\begingroup\let\do\@makeother\dospecials
  \catcode`\{=1 \catcode`\}=2 \doi@aux}
\providecommand{\doi@aux}[1]{\endgroup\texttt{#1}}
\makeatother
\providecommand*\mcitethebibliography{\thebibliography}
\csname @ifundefined\endcsname{endmcitethebibliography}
  {\let\endmcitethebibliography\endthebibliography}{}
\begin{mcitethebibliography}{64}
\providecommand*\natexlab[1]{#1}
\providecommand*\mciteSetBstSublistMode[1]{}
\providecommand*\mciteSetBstMaxWidthForm[2]{}
\providecommand*\mciteBstWouldAddEndPuncttrue
  {\def\EndOfBibitem{\unskip.}}
\providecommand*\mciteBstWouldAddEndPunctfalse
  {\let\EndOfBibitem\relax}
\providecommand*\mciteSetBstMidEndSepPunct[3]{}
\providecommand*\mciteSetBstSublistLabelBeginEnd[3]{}
\providecommand*\EndOfBibitem{}
\mciteSetBstSublistMode{f}
\mciteSetBstMaxWidthForm{subitem}{(\alph{mcitesubitemcount})}
\mciteSetBstSublistLabelBeginEnd
  {\mcitemaxwidthsubitemform\space}
  {\relax}
  {\relax}

\bibitem[Neutze \latin{et~al.}(2000)Neutze, Wouts, van~der Spoel, Weckert, and
  Hajdu]{Neutze:Nature406:752}
Neutze,~R.; Wouts,~R.; van~der Spoel,~D.; Weckert,~E.; Hajdu,~J. Potential for
  biomolecular imaging with femtosecond X-ray pulses. \emph{Nature}
  \textbf{2000}, \emph{406}, 752--757\relax
\mciteBstWouldAddEndPuncttrue
\mciteSetBstMidEndSepPunct{\mcitedefaultmidpunct}
{\mcitedefaultendpunct}{\mcitedefaultseppunct}\relax
\EndOfBibitem
\bibitem[Barty \latin{et~al.}(2013)Barty, K{\"u}pper, and
  Chapman]{Barty:ARPC64:415}
Barty,~A.; K{\"u}pper,~J.; Chapman,~H.~N. Molecular Imaging Using X-Ray
  Free-Electron Lasers. \emph{Annu. Rev. Phys. Chem.} \textbf{2013}, \emph{64},
  415--435\relax
\mciteBstWouldAddEndPuncttrue
\mciteSetBstMidEndSepPunct{\mcitedefaultmidpunct}
{\mcitedefaultendpunct}{\mcitedefaultseppunct}\relax
\EndOfBibitem
\bibitem[Spence and Doak(2004)Spence, and Doak]{Spence:PRL92:198102}
Spence,~J. C.~H.; Doak,~R.~B. Single molecule diffraction. \emph{Phys. Rev.
  Lett.} \textbf{2004}, \emph{92}, 198102\relax
\mciteBstWouldAddEndPuncttrue
\mciteSetBstMidEndSepPunct{\mcitedefaultmidpunct}
{\mcitedefaultendpunct}{\mcitedefaultseppunct}\relax
\EndOfBibitem
\bibitem[Chapman(2009)]{Chapman:NatMater8:299}
Chapman,~H.~N. X-ray imaging beyond the limits. \emph{Nature Mater.}
  \textbf{2009}, \emph{8}, 299--301\relax
\mciteBstWouldAddEndPuncttrue
\mciteSetBstMidEndSepPunct{\mcitedefaultmidpunct}
{\mcitedefaultendpunct}{\mcitedefaultseppunct}\relax
\EndOfBibitem
\bibitem[Bogan \latin{et~al.}(2008)Bogan, Benner, Boutet, Rohner, Frank, Barty,
  Seibert, Maia, Marchesini, Bajt, Woods, Riot, Hau-Riege, Svenda, Marklund,
  Spiller, Hajdu, and Chapman]{Bogan:NanoLett8:310}
Bogan,~M.~J. \latin{et~al.}  Single particle X-ray diffractive imaging.
  \emph{Nano Lett.} \textbf{2008}, \emph{8}, 310--316\relax
\mciteBstWouldAddEndPuncttrue
\mciteSetBstMidEndSepPunct{\mcitedefaultmidpunct}
{\mcitedefaultendpunct}{\mcitedefaultseppunct}\relax
\EndOfBibitem
\bibitem[Spence(2019)]{Spence:SpringerHandbook:SPI:2019}
Spence,~J. C.~H. \emph{Springer Handbook of Microscopy}; Springer Verlag, 2019;
  pp 1009--1036\relax
\mciteBstWouldAddEndPuncttrue
\mciteSetBstMidEndSepPunct{\mcitedefaultmidpunct}
{\mcitedefaultendpunct}{\mcitedefaultseppunct}\relax
\EndOfBibitem
\bibitem[Seibert \latin{et~al.}(2011)Seibert, Ekeberg, Maia, Svenda,
  Andreasson, J{\"o}nsson, Odi{\'c}, Iwan, Rocker, Westphal, Hantke, Deponte,
  Barty, Schulz, Gumprecht, Coppola, Aquila, Liang, White, Martin, Caleman,
  Stern, Abergel, Seltzer, Claverie, Bostedt, Bozek, Boutet, Miahnahri,
  Messerschmidt, Krzywinski, Williams, Hodgson, Bogan, Hampton, Sierra,
  Starodub, Andersson, Bajt, Barthelmess, Spence, Fromme, Weierstall, Kirian,
  Hunter, Doak, Marchesini, Hau-Riege, Frank, Shoeman, Lomb, Epp, Hartmann,
  Rolles, Rudenko, Schmidt, Foucar, Kimmel, Holl, Rudek, Erk, H{\"o}mke, Reich,
  Pietschner, Weidenspointner, Str{\"u}der, Hauser, Gorke, Ullrich,
  Schlichting, Herrmann, Schaller, Schopper, Soltau, K{\"u}hnel, Andritschke,
  Schr{\"o}ter, Krasniqi, Bott, Schorb, Rupp, Adolph, Gorkhover, Hirsemann,
  Potdevin, Graafsma, Nilsson, Chapman, and Hajdu]{Seibert:Nature470:78}
Seibert,~M.~M. \latin{et~al.}  Single mimivirus particles intercepted and
  imaged with an {X}-ray laser. \emph{Nature} \textbf{2011}, \emph{470},
  78--81\relax
\mciteBstWouldAddEndPuncttrue
\mciteSetBstMidEndSepPunct{\mcitedefaultmidpunct}
{\mcitedefaultendpunct}{\mcitedefaultseppunct}\relax
\EndOfBibitem
\bibitem[Ayyer \latin{et~al.}(2021)Ayyer, Xavier, Bielecki, Shen, Daurer,
  Samanta, Awel, Bean, Barty, Bergemann, Ekeberg, Estillore, Fangohr,
  Giewekemeyer, Hunter, Karnevskiy, Kirian, Kirkwood, Kim, Koliyadu, Lange,
  Letrun, L\"{u}bke, Michelat, Morgan, Roth, Sato, Sikorski, Schulz, Spence,
  Vagovic, Wollweber, Worbs, Yefanov, Zhuang, Maia, Horke, K\"{u}pper, Loh,
  Mancuso, and Chapman]{Ayyer:Optica8:15}
Ayyer,~K. \latin{et~al.}  {3D} diffractive imaging of nanoparticle ensembles
  using an X-ray laser. \emph{Optica} \textbf{2021}, \emph{8}, 15--23\relax
\mciteBstWouldAddEndPuncttrue
\mciteSetBstMidEndSepPunct{\mcitedefaultmidpunct}
{\mcitedefaultendpunct}{\mcitedefaultseppunct}\relax
\EndOfBibitem
\bibitem[H\"{o}ing \latin{et~al.}(2023)H\"{o}ing, Salzwedel, Worbs, Zhuang,
  Samanta, L\"{u}bke, Estillore, Dlugolecki, Passow, Erk, Ekanayake, Ramm,
  Correa, Papadopoulou, Noor, Schulz, Selig, Knorr, Ayyer, K\"{u}pper, and
  Lange]{Hoeing:NanoLett23:5943}
H\"{o}ing,~D. \latin{et~al.}  Time-Resolved Single-Particle X-ray Scattering
  Reveals Electron-Density Gradients As Coherent
  Plasmonic-Nanoparticle-Oscillation Source. \emph{Nano Lett.} \textbf{2023},
  \emph{23}, 5943–5950\relax
\mciteBstWouldAddEndPuncttrue
\mciteSetBstMidEndSepPunct{\mcitedefaultmidpunct}
{\mcitedefaultendpunct}{\mcitedefaultseppunct}\relax
\EndOfBibitem
\bibitem[Ayyer \latin{et~al.}(2016)Ayyer, Yefanov, Oberth{\"u}r, Roy-Chowdhury,
  Galli, Mariani, Basu, Coe, Conrad, Fromme, Schaffer, D{\"o}rner, James,
  Kupitz, Metz, Nelson, Xavier, Beyerlein, Schmidt, Sarrou, Spence, Weierstall,
  White, Yang, Zhao, Liang, Aquila, Hunter, Robinson, Koglin, Boutet, Fromme,
  Barty, and Chapman]{Ayyer:Nature530:202}
Ayyer,~K. \latin{et~al.}  Macromolecular diffractive imaging using imperfect
  crystals. \emph{Nature} \textbf{2016}, \emph{530}, 202--206\relax
\mciteBstWouldAddEndPuncttrue
\mciteSetBstMidEndSepPunct{\mcitedefaultmidpunct}
{\mcitedefaultendpunct}{\mcitedefaultseppunct}\relax
\EndOfBibitem
\bibitem[Ayyer(2020)]{Ayyer:Optica7:593}
Ayyer,~K. Reference-enhanced x-ray single-particle imaging. \emph{Optica}
  \textbf{2020}, \emph{7}, 593--601\relax
\mciteBstWouldAddEndPuncttrue
\mciteSetBstMidEndSepPunct{\mcitedefaultmidpunct}
{\mcitedefaultendpunct}{\mcitedefaultseppunct}\relax
\EndOfBibitem
\bibitem[not()]{note:alignment+orientation}
We distiguish molecular alignment and orientation according to the usual
  convention~\cite{Stapelfeldt:RMP75:543}, i.\,e., alignment refers to fixing
  the molecules along certain axes whereas orientation refers to the breaking
  of the ``up-down'' (mirror) symmetry along these axes.\relax
\mciteBstWouldAddEndPunctfalse
\mciteSetBstMidEndSepPunct{\mcitedefaultmidpunct}
{}{\mcitedefaultseppunct}\relax
\EndOfBibitem
\bibitem[Filsinger \latin{et~al.}(2011)Filsinger, Meijer, Stapelfeldt, Chapman,
  and Küpper]{Filsinger:PCCP13:2076}
Filsinger,~F.; Meijer,~G.; Stapelfeldt,~H.; Chapman,~H.~N.; Küpper,~J. State-
  and conformer-selected beams of aligned and oriented molecules for ultrafast
  diffraction studies. \emph{Phys. Chem. Chem. Phys.} \textbf{2011}, \emph{13},
  2076--2087\relax
\mciteBstWouldAddEndPuncttrue
\mciteSetBstMidEndSepPunct{\mcitedefaultmidpunct}
{\mcitedefaultendpunct}{\mcitedefaultseppunct}\relax
\EndOfBibitem
\bibitem[Spence \latin{et~al.}(2005)Spence, Schmidt, Wu, Hembree, Weierstall,
  Doak, and Fromme]{Spence:ActaCrystA61:237}
Spence,~J. C.~H.; Schmidt,~K.; Wu,~J.~S.; Hembree,~G.; Weierstall,~U.;
  Doak,~R.~B.; Fromme,~P. Diffraction and imaging from a beam of laser-aligned
  proteins: resolution limits. \emph{Acta\ Cryst. A} \textbf{2005}, \emph{61},
  237--245\relax
\mciteBstWouldAddEndPuncttrue
\mciteSetBstMidEndSepPunct{\mcitedefaultmidpunct}
{\mcitedefaultendpunct}{\mcitedefaultseppunct}\relax
\EndOfBibitem
\bibitem[Friedrich and Herschbach(1991)Friedrich, and
  Herschbach]{Friedrich:Nature353:412}
Friedrich,~B.; Herschbach,~D.~R. Spatial Orientation of Molecules in Strong
  Electric Fields and Evidence for Pendular States. \emph{Nature}
  \textbf{1991}, \emph{353}, 412--414\relax
\mciteBstWouldAddEndPuncttrue
\mciteSetBstMidEndSepPunct{\mcitedefaultmidpunct}
{\mcitedefaultendpunct}{\mcitedefaultseppunct}\relax
\EndOfBibitem
\bibitem[Block \latin{et~al.}(1992)Block, Bohac, and Miller]{Block:PRL68:1303}
Block,~P.~A.; Bohac,~E.~J.; Miller,~R.~E. Spectroscopy of Pendular States --
  The Use of Molecular Complexes in Achieving Orientation. \emph{Phys. Rev.
  Lett.} \textbf{1992}, \emph{68}, 1303--1306\relax
\mciteBstWouldAddEndPuncttrue
\mciteSetBstMidEndSepPunct{\mcitedefaultmidpunct}
{\mcitedefaultendpunct}{\mcitedefaultseppunct}\relax
\EndOfBibitem
\bibitem[Rosca-Pruna and Vrakking(2001)Rosca-Pruna, and
  Vrakking]{RoscaPruna:PRL87:153902}
Rosca-Pruna,~F.; Vrakking,~M. J.~J. Experimental observation of revival
  structures in picosecond laser-induced alignment of {I}$_2$. \emph{Phys. Rev.
  Lett.} \textbf{2001}, \emph{87}, 153902\relax
\mciteBstWouldAddEndPuncttrue
\mciteSetBstMidEndSepPunct{\mcitedefaultmidpunct}
{\mcitedefaultendpunct}{\mcitedefaultseppunct}\relax
\EndOfBibitem
\bibitem[Larsen \latin{et~al.}(1999)Larsen, Sakai, Safvan, Wendt-Larsen, and
  Stapelfeldt]{Larsen:JCP111:7774}
Larsen,~J.~J.; Sakai,~H.; Safvan,~C.~P.; Wendt-Larsen,~I.; Stapelfeldt,~H.
  Aligning molecules with intense nonresonant laser fields. \emph{J. Chem.
  Phys.} \textbf{1999}, \emph{111}, 7774\relax
\mciteBstWouldAddEndPuncttrue
\mciteSetBstMidEndSepPunct{\mcitedefaultmidpunct}
{\mcitedefaultendpunct}{\mcitedefaultseppunct}\relax
\EndOfBibitem
\bibitem[Stapelfeldt and Seideman(2003)Stapelfeldt, and
  Seideman]{Stapelfeldt:RMP75:543}
Stapelfeldt,~H.; Seideman,~T. Colloquium: Aligning molecules with strong laser
  pulses. \emph{Rev. Mod. Phys.} \textbf{2003}, \emph{75}, 543--557\relax
\mciteBstWouldAddEndPuncttrue
\mciteSetBstMidEndSepPunct{\mcitedefaultmidpunct}
{\mcitedefaultendpunct}{\mcitedefaultseppunct}\relax
\EndOfBibitem
\bibitem[Trippel \latin{et~al.}(2014)Trippel, Mullins, M{\"u}ller, Kienitz,
  Omiste, Stapelfeldt, Gonz{\'a}lez-F{\'e}rez, and
  K{\"u}pper]{Trippel:PRA89:051401R}
Trippel,~S.; Mullins,~T.; M{\"u}ller,~N. L.~M.; Kienitz,~J.~S.; Omiste,~J.~J.;
  Stapelfeldt,~H.; Gonz{\'a}lez-F{\'e}rez,~R.; K{\"u}pper,~J. Strongly driven
  quantum pendulum of the carbonyl sulfide molecule. \emph{Phys. Rev. A}
  \textbf{2014}, \emph{89}, 051401(R)\relax
\mciteBstWouldAddEndPuncttrue
\mciteSetBstMidEndSepPunct{\mcitedefaultmidpunct}
{\mcitedefaultendpunct}{\mcitedefaultseppunct}\relax
\EndOfBibitem
\bibitem[Koch \latin{et~al.}(2019)Koch, Lemeshko, and Sugny]{Koch:RMP91:035005}
Koch,~C.~P.; Lemeshko,~M.; Sugny,~D. Quantum control of molecular rotation.
  \emph{Rev. Mod. Phys.} \textbf{2019}, \emph{91}, 035005\relax
\mciteBstWouldAddEndPuncttrue
\mciteSetBstMidEndSepPunct{\mcitedefaultmidpunct}
{\mcitedefaultendpunct}{\mcitedefaultseppunct}\relax
\EndOfBibitem
\bibitem[Larsen \latin{et~al.}(2000)Larsen, Hald, Bjerre, Stapelfeldt, and
  Seideman]{Larsen:PRL85:2470}
Larsen,~J.~J.; Hald,~K.; Bjerre,~N.; Stapelfeldt,~H.; Seideman,~T. Three
  dimensional alignment of molecules using elliptically polarized laser fields.
  \emph{Phys. Rev. Lett.} \textbf{2000}, \emph{85}, 2470--2473\relax
\mciteBstWouldAddEndPuncttrue
\mciteSetBstMidEndSepPunct{\mcitedefaultmidpunct}
{\mcitedefaultendpunct}{\mcitedefaultseppunct}\relax
\EndOfBibitem
\bibitem[Holmegaard \latin{et~al.}(2010)Holmegaard, Hansen, Kalh{\o}j, Kragh,
  Stapelfeldt, Filsinger, K{\"u}pper, Meijer, Dimitrovski, Abu-samha, Martiny,
  and Madsen]{Holmegaard:NatPhys6:428}
Holmegaard,~L.; Hansen,~J.~L.; Kalh{\o}j,~L.; Kragh,~S.~L.; Stapelfeldt,~H.;
  Filsinger,~F.; K{\"u}pper,~J.; Meijer,~G.; Dimitrovski,~D.; Abu-samha,~M.;
  Martiny,~C. P.~J.; Madsen,~L.~B. Photoelectron angular distributions from
  strong-field ionization of oriented molecules. \emph{Nat. Phys.}
  \textbf{2010}, \emph{6}, 428--432\relax
\mciteBstWouldAddEndPuncttrue
\mciteSetBstMidEndSepPunct{\mcitedefaultmidpunct}
{\mcitedefaultendpunct}{\mcitedefaultseppunct}\relax
\EndOfBibitem
\bibitem[Holmegaard \latin{et~al.}(2009)Holmegaard, Nielsen, Nevo, Stapelfeldt,
  Filsinger, K{\"u}pper, and Meijer]{Holmegaard:PRL102:023001}
Holmegaard,~L.; Nielsen,~J.~H.; Nevo,~I.; Stapelfeldt,~H.; Filsinger,~F.;
  K{\"u}pper,~J.; Meijer,~G. Laser-induced alignment and orientation of
  quantum-state-selected large molecules. \emph{Phys. Rev. Lett.}
  \textbf{2009}, \emph{102}, 023001\relax
\mciteBstWouldAddEndPuncttrue
\mciteSetBstMidEndSepPunct{\mcitedefaultmidpunct}
{\mcitedefaultendpunct}{\mcitedefaultseppunct}\relax
\EndOfBibitem
\bibitem[Trippel \latin{et~al.}(2013)Trippel, Mullins, M{\"u}ller, Kienitz,
  D{\l}ugo{\l}\k{e}cki, and K{\"u}pper]{Trippel:MP111:1738}
Trippel,~S.; Mullins,~T.~G.; M{\"u}ller,~N. L.~M.; Kienitz,~J.~S.;
  D{\l}ugo{\l}\k{e}cki,~K.; K{\"u}pper,~J. Strongly aligned and oriented
  molecular samples at a {kHz} repetition rate. \emph{Mol. Phys.}
  \textbf{2013}, \emph{111}, 1738\relax
\mciteBstWouldAddEndPuncttrue
\mciteSetBstMidEndSepPunct{\mcitedefaultmidpunct}
{\mcitedefaultendpunct}{\mcitedefaultseppunct}\relax
\EndOfBibitem
\bibitem[Karamatskos \latin{et~al.}(2019)Karamatskos, Raabe, Mullins,
  Trabattoni, Stammer, Goldsztejn, Johansen, D{\l}ugo{\l}\k{e}cki, Stapelfeldt,
  Vrakking, Trippel, Rouzée, and Küpper]{Karamatskos:NatComm10:3364}
Karamatskos,~E.~T.; Raabe,~S.; Mullins,~T.; Trabattoni,~A.; Stammer,~P.;
  Goldsztejn,~G.; Johansen,~R.~R.; D{\l}ugo{\l}\k{e}cki,~K.; Stapelfeldt,~H.;
  Vrakking,~M. J.~J.; Trippel,~S.; Rouzée,~A.; Küpper,~J. Molecular movie of
  ultrafast coherent rotational dynamics of {OCS}. \emph{Nat. Commun.}
  \textbf{2019}, \emph{10}, 3364\relax
\mciteBstWouldAddEndPuncttrue
\mciteSetBstMidEndSepPunct{\mcitedefaultmidpunct}
{\mcitedefaultendpunct}{\mcitedefaultseppunct}\relax
\EndOfBibitem
\bibitem[Mullins \latin{et~al.}(2022)Mullins, Karamatskos, Wiese, Onvlee,
  Rouz\'{e}e, Yachmenev, Trippel, and K\"{u}pper]{Mullins:NatComm13:1431}
Mullins,~T.; Karamatskos,~E.~T.; Wiese,~J.; Onvlee,~J.; Rouz\'{e}e,~A.;
  Yachmenev,~A.; Trippel,~S.; K\"{u}pper,~J. Picosecond pulse-shaping for
  strong three-dimensional field-free alignment of generic asymmetric-top
  molecules. \emph{Nat. Commun.} \textbf{2022}, \emph{13}, 1431\relax
\mciteBstWouldAddEndPuncttrue
\mciteSetBstMidEndSepPunct{\mcitedefaultmidpunct}
{\mcitedefaultendpunct}{\mcitedefaultseppunct}\relax
\EndOfBibitem
\bibitem[Pentlehner \latin{et~al.}(2013)Pentlehner, Nielsen, Christiansen,
  Slenczka, and Stapelfeldt]{Pentlehner:PRA87:063401}
Pentlehner,~D.; Nielsen,~J.~H.; Christiansen,~L.; Slenczka,~A.; Stapelfeldt,~H.
  Laser-induced adiabatic alignment of molecules dissolved in helium
  nanodroplets. \emph{Phys. Rev. A} \textbf{2013}, \emph{87}, 063401\relax
\mciteBstWouldAddEndPuncttrue
\mciteSetBstMidEndSepPunct{\mcitedefaultmidpunct}
{\mcitedefaultendpunct}{\mcitedefaultseppunct}\relax
\EndOfBibitem
\bibitem[Chatterley \latin{et~al.}(2019)Chatterley, Schouder, Christiansen,
  Shepperson, Rasmussen, and Stapelfeldt]{Chatterley:NatCommun10:133}
Chatterley,~A.~S.; Schouder,~C.; Christiansen,~L.; Shepperson,~B.;
  Rasmussen,~M.~H.; Stapelfeldt,~H. Long-lasting field-free alignment of large
  molecules inside helium nanodroplets. \emph{Nat. Commun.} \textbf{2019},
  \emph{10}, 133\relax
\mciteBstWouldAddEndPuncttrue
\mciteSetBstMidEndSepPunct{\mcitedefaultmidpunct}
{\mcitedefaultendpunct}{\mcitedefaultseppunct}\relax
\EndOfBibitem
\bibitem[Trippel \latin{et~al.}(2018)Trippel, Wiese, Mullins, and
  K{\"u}pper]{Trippel:JCP148:101103}
Trippel,~S.; Wiese,~J.; Mullins,~T.; K{\"u}pper,~J. Communication: Strong laser
  alignment of solvent-solute aggregates in the gas-phase. \emph{J. Chem.
  Phys.} \textbf{2018}, \emph{148}, 101103\relax
\mciteBstWouldAddEndPuncttrue
\mciteSetBstMidEndSepPunct{\mcitedefaultmidpunct}
{\mcitedefaultendpunct}{\mcitedefaultseppunct}\relax
\EndOfBibitem
\bibitem[Park \latin{et~al.}(2008)Park, Gahlmann, He, Feenstra, and
  Zewail]{Park:ACIE47:9496}
Park,~S.~T.; Gahlmann,~A.; He,~Y.; Feenstra,~J.~S.; Zewail,~A.~H. Ultrafast
  Electron Diffraction Reveals Dark Structures of the Biological Chromophore
  Indole. \emph{Angew. Chem. Int. Ed.} \textbf{2008}, \emph{47},
  9496--9499\relax
\mciteBstWouldAddEndPuncttrue
\mciteSetBstMidEndSepPunct{\mcitedefaultmidpunct}
{\mcitedefaultendpunct}{\mcitedefaultseppunct}\relax
\EndOfBibitem
\bibitem[Hensley \latin{et~al.}(2012)Hensley, Yang, and
  Centurion]{Hensley:PRL109:133202}
Hensley,~C.~J.; Yang,~J.; Centurion,~M. Imaging of Isolated Molecules with
  Ultrafast Electron Pulses. \emph{Phys. Rev. Lett.} \textbf{2012}, \emph{109},
  133202\relax
\mciteBstWouldAddEndPuncttrue
\mciteSetBstMidEndSepPunct{\mcitedefaultmidpunct}
{\mcitedefaultendpunct}{\mcitedefaultseppunct}\relax
\EndOfBibitem
\bibitem[K{\"u}pper \latin{et~al.}(2014)K{\"u}pper, Stern, Holmegaard,
  Filsinger, Rouz\'{e}e, Rudenko, Johnsson, Martin, Adolph, Aquila, Bajt,
  Barty, Bostedt, Bozek, Caleman, Coffee, Coppola, Delmas, Epp, Erk, Foucar,
  Gorkhover, Gumprecht, Hartmann, Hartmann, Hauser, Holl, H{\"o}mke, Kimmel,
  Krasniqi, K{\"u}hnel, Maurer, Messerschmidt, Moshammer, Reich, Rudek, Santra,
  Schlichting, Schmidt, Schorb, Schulz, Soltau, Spence, Starodub, Str{\"u}der,
  Th{\o}gersen, Vrakking, Weidenspointner, White, Wunderer, Meijer, Ullrich,
  Stapelfeldt, Rolles, and Chapman]{Kuepper:PRL112:083002}
K{\"u}pper,~J. \latin{et~al.}  X-Ray Diffraction from Isolated and Strongly
  Aligned Gas-Phase Molecules with a Free-Electron Laser. \emph{Phys. Rev.
  Lett.} \textbf{2014}, \emph{112}, 083002\relax
\mciteBstWouldAddEndPuncttrue
\mciteSetBstMidEndSepPunct{\mcitedefaultmidpunct}
{\mcitedefaultendpunct}{\mcitedefaultseppunct}\relax
\EndOfBibitem
\bibitem[Noé(2015)]{Noe:BPJ108:228}
Noé,~F. Beating the Millisecond Barrier in Molecular Dynamics Simulations.
  \emph{Biophys. J.} \textbf{2015}, \emph{108}, 228--229\relax
\mciteBstWouldAddEndPuncttrue
\mciteSetBstMidEndSepPunct{\mcitedefaultmidpunct}
{\mcitedefaultendpunct}{\mcitedefaultseppunct}\relax
\EndOfBibitem
\bibitem[Marklund \latin{et~al.}(2017)Marklund, Ekeberg, Moog, Benesch, and
  Caleman]{Marklund:JPCL8:4540}
Marklund,~E.~G.; Ekeberg,~T.; Moog,~M.; Benesch,~J. L.~P.; Caleman,~C.
  Controlling Protein Orientation in Vacuum Using Electric Fields. \emph{J.
  Phys. Chem. Lett.} \textbf{2017}, \emph{8}, 4540–4544\relax
\mciteBstWouldAddEndPuncttrue
\mciteSetBstMidEndSepPunct{\mcitedefaultmidpunct}
{\mcitedefaultendpunct}{\mcitedefaultseppunct}\relax
\EndOfBibitem
\bibitem[Brodmerkel \latin{et~al.}(2023)Brodmerkel, Santis, Caleman, and
  Marklund]{Brodmerkel:ProteinJ42:205}
Brodmerkel,~M.~N.; Santis,~E.~D.; Caleman,~C.; Marklund,~E.~G. Rehydration
  Post-orientation: Investigating Field-Induced Structural Changes via
  Computational Rehydration. \emph{Protein J.} \textbf{2023}, \emph{42},
  205–218\relax
\mciteBstWouldAddEndPuncttrue
\mciteSetBstMidEndSepPunct{\mcitedefaultmidpunct}
{\mcitedefaultendpunct}{\mcitedefaultseppunct}\relax
\EndOfBibitem
\bibitem[Juba \latin{et~al.}(2017)Juba, Audus, Mascagni, Douglas, and
  Keyrouz]{Juba:JRNIST122:20}
Juba,~D.; Audus,~D.~J.; Mascagni,~M.; Douglas,~J.~F.; Keyrouz,~W. {ZENO}:
  Software for calculating hydrodynamic, electrical, and shape properties of
  polymer and particle suspensions. \emph{J. Res. Nat. Inst. Stand. Tech.}
  \textbf{2017}, \emph{122}, 20\relax
\mciteBstWouldAddEndPuncttrue
\mciteSetBstMidEndSepPunct{\mcitedefaultmidpunct}
{\mcitedefaultendpunct}{\mcitedefaultseppunct}\relax
\EndOfBibitem
\bibitem[Amin \latin{et~al.}(2019)Amin, Samy, and Küpper]{Amin:JPCL10:2938}
Amin,~M.; Samy,~H.; Küpper,~J. Robust and accurate computational estimation of
  the polarizability tensors of macromolecules. \emph{J. Phys. Chem. Lett.}
  \textbf{2019}, \emph{10}, 2938--2943\relax
\mciteBstWouldAddEndPuncttrue
\mciteSetBstMidEndSepPunct{\mcitedefaultmidpunct}
{\mcitedefaultendpunct}{\mcitedefaultseppunct}\relax
\EndOfBibitem
\bibitem[Amin and Küpper(2020)Amin, and Küpper]{Amin:ChemistryOpen9:691}
Amin,~M.; Küpper,~J. Variations in Proteins Dielectric Constants.
  \emph{ChemistryOpen} \textbf{2020}, \emph{9}, 691–69\relax
\mciteBstWouldAddEndPuncttrue
\mciteSetBstMidEndSepPunct{\mcitedefaultmidpunct}
{\mcitedefaultendpunct}{\mcitedefaultseppunct}\relax
\EndOfBibitem
\bibitem[Ma \latin{et~al.}(2021)Ma, Coudert, Billard, Bournazel, Lavorel, Wu,
  Maroulis, Hartmann, and Faucher]{Ma:PRR3:023192}
Ma,~J.; Coudert,~L.~H.; Billard,~F.; Bournazel,~M.; Lavorel,~B.; Wu,~J.;
  Maroulis,~G.; Hartmann,~J.-M.; Faucher,~O. Echo-assisted impulsive alignment
  of room-temperature acetone molecules. \emph{Phys. Rev. Research}
  \textbf{2021}, \emph{3}, 023192\relax
\mciteBstWouldAddEndPuncttrue
\mciteSetBstMidEndSepPunct{\mcitedefaultmidpunct}
{\mcitedefaultendpunct}{\mcitedefaultseppunct}\relax
\EndOfBibitem
\bibitem[Berman \latin{et~al.}(2000)Berman, Westbrook, Feng, Gilliland, Bhat,
  Weissig, Shindyalov, and Bourne]{Berman:NuclAcidRes28:235}
Berman,~H.~M.; Westbrook,~J.; Feng,~Z.; Gilliland,~G.; Bhat,~T.~N.;
  Weissig,~H.; Shindyalov,~I.~N.; Bourne,~P.~E. The {Protein} {Data} {Bank}.
  \emph{Nucleic Acids Research} \textbf{2000}, \emph{28}, 235--242\relax
\mciteBstWouldAddEndPuncttrue
\mciteSetBstMidEndSepPunct{\mcitedefaultmidpunct}
{\mcitedefaultendpunct}{\mcitedefaultseppunct}\relax
\EndOfBibitem
\bibitem[Amin and Küpper(2023)Amin, and Küpper]{Amin:CMIclassirot}
Amin,~M.; Küpper,~J. {CMIclassirot}: Classical-physics simulations of laser
  alignment, \url{https://github.com/CFEL-CMI/CMIclassirot.git}. Software
  package, 2023; \url{https://github.com/CFEL-CMI/CMIclassirot.git}\relax
\mciteBstWouldAddEndPuncttrue
\mciteSetBstMidEndSepPunct{\mcitedefaultmidpunct}
{\mcitedefaultendpunct}{\mcitedefaultseppunct}\relax
\EndOfBibitem
\bibitem[Hartmann and Boulet(2012)Hartmann, and Boulet]{Hartmann:JCP136:184302}
Hartmann,~J.~M.; Boulet,~C. Quantum and classical approaches for rotational
  relaxation and nonresonant laser alignment of linear molecules: A comparison
  for CO2 gas in the nonadiabatic regime. \emph{J. Chem. Phys.} \textbf{2012},
  \emph{136}, 184302\relax
\mciteBstWouldAddEndPuncttrue
\mciteSetBstMidEndSepPunct{\mcitedefaultmidpunct}
{\mcitedefaultendpunct}{\mcitedefaultseppunct}\relax
\EndOfBibitem
\bibitem[Kooij and B.(2006)Kooij, and B.]{Kooij:PCCP8:3349}
Kooij,~E.~S.; B.,~P. Shape and size effects in the optical properties of
  metallic nanorods. \emph{Phys. Chem. Chem. Phys.} \textbf{2006}, \emph{8},
  3349--3357\relax
\mciteBstWouldAddEndPuncttrue
\mciteSetBstMidEndSepPunct{\mcitedefaultmidpunct}
{\mcitedefaultendpunct}{\mcitedefaultseppunct}\relax
\EndOfBibitem
\bibitem[Samanta \latin{et~al.}(2020)Samanta, Amin, Estillore, Roth, Worbs,
  Horke, and Küpper]{Samanta:StructDyn7:024304}
Samanta,~A.~K.; Amin,~M.; Estillore,~A.~D.; Roth,~N.; Worbs,~L.; Horke,~D.~A.;
  Küpper,~J. Controlled beams of shockfrozen, isolated, biological and
  artificial nanoparticles. \emph{Struct. Dyn.} \textbf{2020}, \emph{7},
  024304\relax
\mciteBstWouldAddEndPuncttrue
\mciteSetBstMidEndSepPunct{\mcitedefaultmidpunct}
{\mcitedefaultendpunct}{\mcitedefaultseppunct}\relax
\EndOfBibitem
\bibitem[Kumarappan \latin{et~al.}(2006)Kumarappan, Bisgaard, Viftrup,
  Holmegaard, and Stapelfeldt]{Kumarappan:JCP125:194309}
Kumarappan,~V.; Bisgaard,~C.~Z.; Viftrup,~S.~S.; Holmegaard,~L.;
  Stapelfeldt,~H. Role of rotational temperature in adiabatic molecular
  alignment. \emph{J. Chem. Phys.} \textbf{2006}, \emph{125}, 194309\relax
\mciteBstWouldAddEndPuncttrue
\mciteSetBstMidEndSepPunct{\mcitedefaultmidpunct}
{\mcitedefaultendpunct}{\mcitedefaultseppunct}\relax
\EndOfBibitem
\bibitem[Yang \latin{et~al.}(1996)Yang, Moss, and
  Phillips~Jr.]{Yang:NatBio14:1246}
Yang,~F.; Moss,~L.~G.; Phillips~Jr.,~G.~N. The molecular structure of green
  fluorescent protein. \emph{Nat. Biotechnol.} \textbf{1996}, \emph{14},
  1246--1251\relax
\mciteBstWouldAddEndPuncttrue
\mciteSetBstMidEndSepPunct{\mcitedefaultmidpunct}
{\mcitedefaultendpunct}{\mcitedefaultseppunct}\relax
\EndOfBibitem
\bibitem[Hartmann \latin{et~al.}(1996)Hartmann, Miller, Toennies, and
  Vilesov]{Hartmann:Science272:1631}
Hartmann,~M.; Miller,~R.~E.; Toennies,~J.~P.; Vilesov,~A.~F. High-Resolution
  Molecular Spectroscopy of van der Waals Clusters in Liquid Helium Droplets.
  \emph{Science} \textbf{1996}, \emph{272}, 1631--1634\relax
\mciteBstWouldAddEndPuncttrue
\mciteSetBstMidEndSepPunct{\mcitedefaultmidpunct}
{\mcitedefaultendpunct}{\mcitedefaultseppunct}\relax
\EndOfBibitem
\bibitem[Choi \latin{et~al.}(2006)Choi, Douberly, Falconer, Lewis, Lindsay,
  Merritt, Stiles, and Miller]{Choi:IRPC25:15}
Choi,~M.~Y.; Douberly,~G.~E.; Falconer,~T.~M.; Lewis,~W.~K.; Lindsay,~C.~M.;
  Merritt,~J.~M.; Stiles,~P.~L.; Miller,~R.~E. Infrared spectroscopy of helium
  nanodroplets: novel methods for physics and chemistry. \emph{Int. Rev. Phys.
  Chem.} \textbf{2006}, \emph{25}, 15--75\relax
\mciteBstWouldAddEndPuncttrue
\mciteSetBstMidEndSepPunct{\mcitedefaultmidpunct}
{\mcitedefaultendpunct}{\mcitedefaultseppunct}\relax
\EndOfBibitem
\bibitem[Bierau \latin{et~al.}(2010)Bierau, Kupser, Meijer, and von
  Helden]{Bierau:PRL105:133402}
Bierau,~F.; Kupser,~P.; Meijer,~G.; von Helden,~G. Catching Proteins in Liquid
  Helium Droplets. \emph{Phys. Rev. Lett.} \textbf{2010}, \emph{105},
  133402\relax
\mciteBstWouldAddEndPuncttrue
\mciteSetBstMidEndSepPunct{\mcitedefaultmidpunct}
{\mcitedefaultendpunct}{\mcitedefaultseppunct}\relax
\EndOfBibitem
\bibitem[Alghamdi \latin{et~al.}(2017)Alghamdi, Zhang, Oswalt, Porter, Mehl,
  and Kong]{Alghamdi:JPCA121:6671}
Alghamdi,~M.; Zhang,~J.; Oswalt,~A.; Porter,~J.~J.; Mehl,~R.~A.; Kong,~W.
  Doping of Green Fluorescent Protein into Superfluid Helium Droplets: Size and
  Velocity of Doped Droplets. \emph{J. Phys. Chem. A} \textbf{2017},
  \emph{121}, 6671--6678\relax
\mciteBstWouldAddEndPuncttrue
\mciteSetBstMidEndSepPunct{\mcitedefaultmidpunct}
{\mcitedefaultendpunct}{\mcitedefaultseppunct}\relax
\EndOfBibitem
\bibitem[Gomez \latin{et~al.}(2014)Gomez, Ferguson, Cryan, Bacellar, Tanyag,
  Jones, Schorb, Anielski, Belkacem, Bernando, Boll, Bozek, Carron, Chen,
  Delmas, Englert, Epp, Erk, Foucar, Hartmann, Hexemer, Huth, Kwok, Leone, Ma,
  Maia, Malmerberg, Marchesini, Neumark, Poon, Prell, Rolles, Rudek, Rudenko,
  Seifrid, Siefermann, Sturm, Swiggers, Ullrich, Weise, Zwart, Bostedt,
  Gessner, and Vilesov]{Gomez:Science345:906}
Gomez,~L.~F. \latin{et~al.}  Shapes and vorticities of superfluid helium
  nanodroplets. \emph{Science} \textbf{2014}, \emph{345}, 906--909\relax
\mciteBstWouldAddEndPuncttrue
\mciteSetBstMidEndSepPunct{\mcitedefaultmidpunct}
{\mcitedefaultendpunct}{\mcitedefaultseppunct}\relax
\EndOfBibitem
\bibitem[Zhang \latin{et~al.}(2014)Zhang, He, Freund, and
  Kong]{Zhang:JPCL5:1801}
Zhang,~J.; He,~Y.; Freund,~W.~M.; Kong,~W. Electron Diffraction of Superfluid
  Helium Droplets. \emph{J. Phys. Chem. Lett.} \textbf{2014}, \emph{5},
  1801--1805\relax
\mciteBstWouldAddEndPuncttrue
\mciteSetBstMidEndSepPunct{\mcitedefaultmidpunct}
{\mcitedefaultendpunct}{\mcitedefaultseppunct}\relax
\EndOfBibitem
\bibitem[Hutzler \latin{et~al.}(2012)Hutzler, Lu, and
  Doyle]{Hutzler:CR112:4803}
Hutzler,~N.~R.; Lu,~H.-I.; Doyle,~J.~M. The buffer gas beam: An intense, cold,
  and slow source for atoms and molecules. \emph{Chem. Rev.} \textbf{2012},
  \emph{112}, 4803--4827\relax
\mciteBstWouldAddEndPuncttrue
\mciteSetBstMidEndSepPunct{\mcitedefaultmidpunct}
{\mcitedefaultendpunct}{\mcitedefaultseppunct}\relax
\EndOfBibitem
\bibitem[Weinstein \latin{et~al.}(1998)Weinstein, {d}e{C}arvalho, Guillet,
  Friedrich, and Doyle]{Weinstein:Nature395:148}
Weinstein,~J.~D.; {d}e{C}arvalho,~R.; Guillet,~T.; Friedrich,~B.; Doyle,~J.~M.
  Magnetic trapping of calcium monohydride molecules at millikelvin
  temperatures. \emph{Nature} \textbf{1998}, \emph{395}, 148--150\relax
\mciteBstWouldAddEndPuncttrue
\mciteSetBstMidEndSepPunct{\mcitedefaultmidpunct}
{\mcitedefaultendpunct}{\mcitedefaultseppunct}\relax
\EndOfBibitem
\bibitem[Deppe \latin{et~al.}(2015)Deppe, Huber, Kr\"ankel, and
  K\"upper]{Deppe:OptExp23:28491}
Deppe,~B.; Huber,~G.; Kr\"ankel,~C.; K\"upper,~J. High-intracavity-power
  thin-disk laser for alignment of molecules. \emph{Opt. Express}
  \textbf{2015}, \emph{23}, 28491\relax
\mciteBstWouldAddEndPuncttrue
\mciteSetBstMidEndSepPunct{\mcitedefaultmidpunct}
{\mcitedefaultendpunct}{\mcitedefaultseppunct}\relax
\EndOfBibitem
\bibitem[Rischel \latin{et~al.}(1998)Rischel, Spiedel, Ridge, Jones, Breton,
  Lambry, Martin, and Vos]{Rischel:PNAS95:12306}
Rischel,~C.; Spiedel,~D.; Ridge,~J.~P.; Jones,~M.~R.; Breton,~J.;
  Lambry,~J.-C.; Martin,~J.-L.; Vos,~M.~H. Low frequency vibrational modes in
  proteins: Changes induced by point-mutations in the protein-cofactor matrix
  of bacterial reaction centers. \emph{PNAS} \textbf{1998}, \emph{95},
  12306--12311, Lowest vibrational-frequency modes at $\sim0.05$~THz, i.e.,
  1.5 cm-1. Lowest response of protein to perturbation at $\sim0.01$~THz,
  i.e., $>0.1$~cm-1.\relax
\mciteBstWouldAddEndPunctfalse
\mciteSetBstMidEndSepPunct{\mcitedefaultmidpunct}
{}{\mcitedefaultseppunct}\relax
\EndOfBibitem
\bibitem[Ford(2003)]{Ford:PRE67:051924}
Ford,~L.~H. Estimate of the vibrational frequencies of spherical virus
  particles. \emph{Phys. Rev. E} \textbf{2003}, \emph{67}, 051924\relax
\mciteBstWouldAddEndPuncttrue
\mciteSetBstMidEndSepPunct{\mcitedefaultmidpunct}
{\mcitedefaultendpunct}{\mcitedefaultseppunct}\relax
\EndOfBibitem
\bibitem[Bocko and Lengvarský(2014)Bocko, and Lengvarský]{Bocko:ProcEng96:21}
Bocko,~J.; Lengvarský,~P. Bending Vibrations of Carbon Nanotubes by Using
  Nonlocal Theory. \emph{Proc. Eng.} \textbf{2014}, \emph{96}, 21--27\relax
\mciteBstWouldAddEndPuncttrue
\mciteSetBstMidEndSepPunct{\mcitedefaultmidpunct}
{\mcitedefaultendpunct}{\mcitedefaultseppunct}\relax
\EndOfBibitem
\bibitem[Sinelnikova \latin{et~al.}(2021)Sinelnikova, Mandl, Agelii, Grånäs,
  Marklund, Caleman, and De~Santis]{Sinelnikova:BPJ120:3709}
Sinelnikova,~A.; Mandl,~T.; Agelii,~H.; Grånäs,~O.; Marklund,~E.~G.;
  Caleman,~C.; De~Santis,~E. Protein orientation in time-dependent electric
  fields: orientation before destruction. \emph{Biophys. J.} \textbf{2021},
  \emph{120}, 3709--3717\relax
\mciteBstWouldAddEndPuncttrue
\mciteSetBstMidEndSepPunct{\mcitedefaultmidpunct}
{\mcitedefaultendpunct}{\mcitedefaultseppunct}\relax
\EndOfBibitem
\bibitem[Maugeria \latin{et~al.}(2018)Maugeria, Griesed, Brancae, Millerb,
  Smithc, Eiriche, Högbom, and Shafaat]{Maugeria:JACS140:1471}
Maugeria,~P.~T.; Griesed,~J.~J.; Brancae,~R.~M.; Millerb,~E.~K.; Smithc,~Z.~R.;
  Eiriche,~J.; Högbom,~M.; Shafaat,~H.~S. Driving protein conformational
  changes with light: Photoinduced structural rearrangement in a
  heterobimetallic oxidase. \emph{J. Am. Chem. Soc.} \textbf{2018}, \emph{140},
  1471\relax
\mciteBstWouldAddEndPuncttrue
\mciteSetBstMidEndSepPunct{\mcitedefaultmidpunct}
{\mcitedefaultendpunct}{\mcitedefaultseppunct}\relax
\EndOfBibitem
\bibitem[Brinck \latin{et~al.}(1993)Brinck, Murray, and
  Politzer]{Brinck:JCP98:4305}
Brinck,~T.; Murray,~J.; Politzer,~P. Polarizability and volume. \emph{J. Chem.
  Phys.} \textbf{1993}, \emph{98}, 4305--4306\relax
\mciteBstWouldAddEndPuncttrue
\mciteSetBstMidEndSepPunct{\mcitedefaultmidpunct}
{\mcitedefaultendpunct}{\mcitedefaultseppunct}\relax
\EndOfBibitem
\bibitem[Krajewski(2017)]{Krajewski:Nanoscale:9:16511}
Krajewski,~M. Magnetic-field-induced synthesis of magnetic wire-like micro- and
  nanostructures. \emph{Nanoscale} \textbf{2017}, \emph{9}, 16511--16545\relax
\mciteBstWouldAddEndPuncttrue
\mciteSetBstMidEndSepPunct{\mcitedefaultmidpunct}
{\mcitedefaultendpunct}{\mcitedefaultseppunct}\relax
\EndOfBibitem
\bibitem[Stickler \latin{et~al.}(2021)Stickler, Hornberger, and
  Kim]{Stickler:NatRevPhys3:589}
Stickler,~B.~A.; Hornberger,~K.; Kim,~M.~S. Quantum rotations of nanoparticles.
  \emph{Nat. Rev. Phys.} \textbf{2021}, \emph{3}, 589--597\relax
\mciteBstWouldAddEndPuncttrue
\mciteSetBstMidEndSepPunct{\mcitedefaultmidpunct}
{\mcitedefaultendpunct}{\mcitedefaultseppunct}\relax
\EndOfBibitem
\end{mcitethebibliography}
\bibliographystyle{achemso}

\onecolumngrid
\listofnotes
\end{document}